\documentclass[12pt,twoside]{article}

\date{19-2-2012} 

\begin{document} 

\centerline{\bf Adv. Studies Theor. Phys.} 

\centerline{} 

\centerline{} 

\centerline {\Large{\bf Bell's correlation for continuous local hidden causality is ambiguous.}} 

\centerline{} 


\centerline{} 

\centerline{\bf {J.F. Geurdes}} 

\centerline{} 

\centerline{C. vd. Lijnstraat 164 2593 NN Den Haag Netherlands} 

\centerline{han.geurdes@gmail.com} 



\centerline{} 

\begin{abstract} In this letter it is demonstrated that Bell's correlation can be ambiguous for continuously distributed local hidden variables. 
\end{abstract} 


\section{Introduction and model} 

Bell's theorem is based on a general expression for a quantum correlation between two measurements \cite{Bell}
\begin{equation}\label{1}
P(\vec{a},\vec{b})=\int_{\lambda \in \Lambda} \rho_{\lambda} A_{\lambda}(\vec{a})B_{\lambda}(\vec{b})
\end{equation}
Here, $\Lambda$ is the set of all local hidden variables, $\lambda$ and $\rho_{\lambda}$ the probability density thereof. The $A_{\lambda}(\vec{a})$ and $B_{\lambda}(\vec{b})$ are the measurement responses of the instruments A and B. Here, $\vec{a}$ and $\vec{b}$ are unit-length 3-dim parameter vectors and are called the settings of the instruments A and B.

When modelling a spin measurement the measurement functions $A_{\lambda}(\vec{a})$ and $B_{\lambda}(\vec{b})$ are supposed to project in the set $\{-1,1\}$. This is a convenient way to ensure that $-1 \leq P(\vec{a},\vec{b}) \leq 1$. When this set of defining properties are chosen they are fixed in the sense of the discours of science but it must be acknowledged that the selection $A_{\lambda}(\vec{a}) \in \{-1,1\}$ and $B_{\lambda}(\vec{b}) \in \{-1,1\}$ is arbitrary. 

In addition concerning the hidden variables $\lambda \in \Lambda$ we assume they are continuous. Hence, $\rho_{\lambda}$ is a probability density function defined on the set $\Lambda$. Note that when a pair of settings is selected by the experimenter in the experiment $\rho_{\lambda}$ remains unchanged because its defining set is $\Lambda$. Any possible restriction on the liberty of $\lambda$ variables in subsets of $\Lambda$  arising from setting instrument A and instrument B will {\it not} affect the defining set $\Lambda$ and hence the $\rho_{\lambda}$. Scientists who believe that densities change because of settings that could possibly restrict the freedom of the $\lambda$ variables via the behavior of the measurement functions in subsets of $\Lambda$ such as in \cite{JFG1}, are mistaken{\footnote {They forget the workings of the measurement functions that may, locally, define those sets}}.

In the case of a continuous density like for x in the real numbers we may take a Gaussian density
\begin{equation}
\varphi(x)=\frac{1}{\sqrt{2\pi}}e^{-\frac{1}{2} x^2}
\end{equation}
Let us employ this density in our subsequent example. In addition it is easy to see that 
\begin{equation}
\Phi(x)=\int_{-\infty}^{x} \varphi(x')dx'
\end{equation}
Subsequently it is always allowed take: $\rho_{\lambda}=\varphi(x)$ and $\lambda=x$ and we have 
\begin{equation}\label{1a}
\varphi(x)=\frac{d}{dx}\Phi(x).
\end{equation}

Moreover, we model a measurement function with a parameter in the real numbers. In the first place let us start with defining a Heaviside limit
\begin{equation}\label{1aa}
H(x)=\lim_{n\rightarrow \infty}exp\left[-\frac{e^{-nx}}{n}\right]
\end{equation}
This 'Heaviside function' is the basis for a spin measurement modelling function. We see from the previous equation that $H(x)=1$ for $x\geq 0$ and $H(x)=0$ for $x < 0$. Hence, a sign function $sign(x)$ can be defined as in
\begin{equation}\label{1bb}
sign(x)=2H(x)-1
\end{equation}
It can easily be verified that
\begin{equation}\label{1b}
\frac{d}{dx} sign(x) = 2 \delta(x)
\end{equation}
with $\delta(x)$ the Dirac delta \cite{Light}.

If we e.g. inspect the integral
\begin{equation}
K=\int_{-\infty}^{+\infty} \varphi(x) \left(\frac{d}{dx} H(x-a)\right) dx
\end{equation}
it follows straightforwardly that $K=\varphi(a)$ which is the effect of integrating with $\delta(x-a)$. Hence, the expression in (\ref{1b}) using $sign(x)=2H(x)-1$. 

As we can see, the 'function' $sign(x)$ as defined previously behaves as can be expected from a measurement function. Its most accute property is that it projects in $\{-1,1\}$. This being the case, as can be verified rather easily, let us inspect a simplified model of how the use of this type of function will work in the practice of computing the quantum correlation (\ref{1}). For a model system let us take a look at the following integration
\begin{equation}\label{2}
I=\int_{-\infty}^{+\infty} \varphi(x) sign(x-a) dx
\end{equation}
For completeness: the $\varphi(x)$ stands for the LHV probability density and the $sign(x-a)$ acts as measurement function. The $a$ is, like x, a real number. It is not hard to imagine that this type of integration will occur in the evaluation of an LHV based correlation. Note that changing the number $a$ will not affect $\varphi(x)$. Note also that $sign(x-a)$ may define two sets, those x where $x<a$ and those x where $x \geq a$.

Because, 
\begin{equation}\label{3}
sign(x-a)=\frac{1}{sign(x-a)}
\end{equation}
it is easy to acknowledge that perhaps in a continuous integration there could arise an ambiguity in the value of the LHV model of the quantum correlation as generally defined in (\ref{1}). Because generally the Heaviside and Dirac functions are used as differentiable functions in integrations it is pointless to argue against the use of (\ref{3}) in integration (\ref{2}) and subsequent possible differentiation in the evaluation of the integral.

Let us first start from  (\ref{2}) and write using (\ref{1a})
\begin{equation}
I=\int_{-\infty}^{+\infty}  sign(x-a) \frac{d}{dx} \Phi(x) dx
\end{equation}
This integral can be rewritten using partial integration (the elementary mathematical toolkit of Bell's theorem) as
\begin{equation}\label{4}
I=1-\int_{-\infty}^{+\infty}  \Phi(x) \frac{d}{dx}sign(x-a)  dx
\end{equation}
The unity in the integral can be explained with $\left[\Phi(x)sign(x-a)\right]_{-\infty}^{+\infty}=1$ assuming that for $x\rightarrow \infty$ we will always 'leave $a$ behind' because the number $a$ is fixed real. 

Using (\ref{1b}) it follows that
\begin{equation}\label{4a}
I=1-2\Phi(a)
\end{equation}
Subsequently let us employ (\ref{3}) and note that the right hand of (\ref{3}) exactly represents the same series of plus and minus one as the left hand because $-1=\frac{1}{-1}$. Hence we may also write 
\begin{equation}\label{5}
J=\int_{-\infty}^{+\infty} \frac{1}{sign(x-a)} \frac{d}{dx} \Phi(x) dx
\end{equation}
and expect, beacuse of (\ref{3}), that $J=I$. 

Employing partial integration on (\ref{5}) we obtain
\begin{equation}\label{6}
J=1-\int_{-\infty}^{+\infty}  \Phi(x) \frac{d}{dx}\left(\frac{1}{sign(x-a)}\right)  dx
\end{equation}
Noting that
\begin{equation}\label{7}
\frac{d}{dx}\left(\frac{1}{sign(x-a)}\right)= - \frac{1}{sign^2(x-a)}\frac{d}{dx}sign(x-a)
\end{equation}
and ${sign^2(x-a)}=1$ it follows that 
\begin{equation}\label{8}
J=1+2\int_{-\infty}^{+\infty}  \Phi(x) \delta(x-a)  dx
\end{equation}
Hence, 
\begin{equation}\label{9}
J=1+2\Phi(a)
\end{equation}

\section{Conclusion}
Comparing (\ref{9}) with (\ref{4a}) it follows that, generally speaking, $J \neq I$ despite the fact that (\ref{3}) suggests otherwise. This means that Bell's correlation expression for LHV variables hides an ambiguity for continuous local hidden variables and is unfit to use for those type of hidden variables when a suitable measurement function such as in (\ref{1bb}) is used. There is no theoretical ground whatsoever to exclude this type of measurement function from consideration.

The present result is in accordance with the result obtained by Joy Christian \cite{Joy} starting from a topological analysis of the measurement functions in the correlations and also connects to an earlier study of the author \cite{JFG1}. Note that in constructive analysis \cite{Bish} the limit in (\ref{1aa}) {\it is} the definition of the generalised function. Because in physics a constructivistic computer proof of locality and causality is considered the ultimate proof \cite{Gill}, the present result is very relevant to proofs of LHV models. This type of proof is of constructivistic kind and principles of constructivism are implicitly accepted therewith. If a programmer is instructed to approximate the correlation of Bell with the use of  numerical integration the ambiguity will show in the selection of measurement function leading to $I\neq J$.

The conclusion is that Bell's expression for the correlation in (\ref{1}) is ambiguous for continuous local hidden variables and the inequalities derived thereof are unfit to eliminate all local hidden variable explanations either using experimental measurements or with the use of computer program simulations.

{\bf Received: Month 02, 2012}

\end{document}